\documentclass[12pt]{article}
\usepackage{times}
\usepackage{geometry}
\usepackage{aas_macros}
\usepackage{color}
\usepackage{comment}
\usepackage[utf8]{inputenc}
\usepackage[utf8]{inputenc}
\usepackage{enumitem,amssymb}
\usepackage{ragged2e}
\newlist{thematic}{itemize}{8}
\setlist[thematic]{label=$\square$}
\usepackage{pifont}
\usepackage{cite}
\usepackage{hyperref}
\usepackage{soul}
\usepackage{multicol}

\newcommand{\Geff}{G_{\rm eff}}

\begin{document}
\pagenumbering{gobble}

\raggedright
\huge
Astro2020 Science White Paper \linebreak

Dark Energy and Modified Gravity \linebreak
\normalsize

\noindent \textbf{Thematic Areas:}  Cosmology and Fundamental Physics \linebreak
  
\textbf{Principal Author:}

Name: An\v{z}e Slosar	
 \linebreak						
Institution: Brookhaven National Laboratory
 \linebreak
Email: \texttt{anze@bnl.gov}
 \linebreak
Phone:  (631) 344 8012
 \linebreak
 
\textbf{Co-authors:} See next page.
\medskip
 
\textbf{Endorsing Collaborations:} The surveys designed to carry out dark energy and cosmology investigations in the next decade have attracted large collaborations, organized to facilitate many diverse science goals and broad participation within unified data sets.  The following collaborations involving over a thousand people are co-signing this white paper to endorse its content, having reviewed it following their respective internal management processes:\\
$\bullet$ The Dark Energy Spectroscopic Instrument Collaboration (DESI),  \\
$\bullet$ The Euclid Consortium, \\
$\bullet$ The LSST Dark Energy Science Collaboration (LSST-DESC), \\
$\bullet$ The Simons Observatory Collaboration (SO),  \\
$\bullet$ The WFIRST Cosmology Science Investigation Teams 
\medskip

\textbf{Abstract:}

Despite two decades of tremendous experimental and theoretical progress, the riddle of the accelerated expansion of the Universe remains to be solved. On the experimental side, our understanding of the possibilities and limitations of the major dark energy probes has evolved; here we summarize the major probes and their crucial challenges.  On the theoretical side, the taxonomy of explanations for the accelerated expansion rate is better understood, providing clear guidance to the relevant observables. We argue that: i) improving statistical precision and systematic control by taking more data, supporting research efforts to address crucial challenges for each probe, using complementary methods, and relying on cross-correlations is well motivated; ii) blinding of analyses is difficult but ever more important; iii) studies of dark energy and modified gravity are related; and iv) it is crucial that R\&D for a vibrant dark energy program in the 2030s be started now by supporting studies and technical R\&D that will allow embryonic proposals to mature.  Understanding dark energy, arguably the biggest unsolved mystery in both fundamental particle physics and cosmology, will remain one of the focal points of cosmology in the forthcoming decade.

\pagebreak


\newcommand{\Amherst}{University of Massachusetts, Amherst, MA 01003 USA}
\newcommand{\ANLHEP}{HEP Division, Argonne National Laboratory, Lemont, IL 60439, USA}
\newcommand{\APC}{Laboratoire Astroparticule et Cosmologie (APC), CNRS/IN2P3, Universit\'e Paris Diderot, 10, rue Alice Domon et Léonie Duquet, 75205 Paris Cedex 13, France}
\newcommand{\ASU}{Arizona State University, Tempe, AZ  85287}
\newcommand{\BenGurion}{Department of Physics, Ben-Gurion University, Be'er Sheva 84105, Israel}
\newcommand{\BNL}{Brookhaven National Laboratory, Upton, NY 11973}
\newcommand{\Brown}{Brown University, Providence, RI 02912}
\newcommand{\Bub}{Boston University, Boston, MA 02215}
\newcommand{\BU}{Boston University, Boston, MA 02215}
\newcommand{\Buffalo}{Department of Physics, University at Buffalo, SUNY Buffalo, NY 14260 USA}
\newcommand{\Caltech}{California Institute of Technology, Pasadena, CA 91125}
\newcommand{\Cardiff}{School of Physics and Astronomy, Cardiff University, The Parade, Cardiff, CF24 3AA, UK}
\newcommand{\Carleton}{Carleton University, K1S 5B6 Ottawa, Canada}
\newcommand{\Carnegie}{The Observatories of the Carnegie Institution for Science, 813 Santa Barbara St., Pasadena, CA 91101, USA}
\newcommand{\Cavendish}{Astrophysics Group, Cavendish Laboratory, J.J.Thomson Avenue, Cambridge, CB3 0HE, UK}
\newcommand{\CCA}{Center for Computational Astrophysics, 162 5th Ave, 10010, New York, NY, USA}
\newcommand{\CPPM}{Aix Marseille Univ, CNRS/IN2P3, CPPM, Marseille, France}
\newcommand{\CEADAP}{D\'epartement d’Astrophysique, CEA Saclay DSM/Irfu, 91191 Gif-sur-Yvette, France}
\newcommand{\CERN}{CERN, Geneva, Switzerland}
\newcommand{\CfA}{Harvard-Smithsonian Center for Astrophysics, MA 02138}
\newcommand{\CFT}{Center for Theoretical Physics, Polish Academy of Sciences, al. Lotnik\'{o}w 32/46, 02-668, Warsaw, Poland}
\newcommand{\Cincinnati}{University of Cincinnati, Cincinnati, OH 45221}
\newcommand{\CITA}{Canadian Institute for Theoretical Astrophysics, University of Toronto, Toronto, ON M5S 3H8, Canada}
\newcommand{\CNRSA}{CNRS, Laboratoire d'Annecy-le-Vieux de Physique Th\'{e}orique, France}
\newcommand{\CNYang}{C.N. Yang Institute for Theoretical Physics State University of New York Stony Brook, NY 11794}
\newcommand{\CMUCosmo}{Department 
of Physics, McWilliams Center for Cosmology, Carnegie Mellon University}
\newcommand{\Columbia}{Columbia University, New York, NY 10027}
\newcommand{\Cornell}{Cornell University, Ithaca, NY 14853}
\newcommand{\CPthree}{CP3-Origins, 5230 Odense, Denmark}
\newcommand{\CUBoulder}{Center for Astrophysics and Space Astronomy, University of Colorado, Boulder,CO 80309, USA}
\newcommand{\CWRU}{Case Western Reserve University, Cleveland, OH 44106}
\newcommand{\daa}{Department of Astronomy and Astrophysics, University of Toronto, ON, M5S3H4}
\newcommand{\damtp}{DAMTP, Centre for Mathematical Sciences, Wilberforce Road, Cambridge, UK, CB3 0WA}
\newcommand{\DESY}{DESY,  22607 Hamburg, Germany}
\newcommand{\DFI}{Departamento de F\'isica, FCFM, Universidad de Chile, Blanco Encalada 2008, Santiago, Chile}
\newcommand{\DOE}{US. Department of Energy, Germantown, MD 20874}
\newcommand{\drexel}{Drexel University, Philadelphia, PA 19104}
\newcommand{\Duke}{Duke University and Triangle Universitites Nuclear Laboratory, Durham, NC 27708}
\newcommand{\DukePhys}{Department of Physics, Duke University, Durham, NC 27708, USA}
\newcommand{\dunlap}{Dunlap Institute for Astronomy and Astrophysics, University of Toronto, ON, M5S3H4}
\newcommand{\Durham}{Department of Physics, Lower Mountjoy, South Rd, Durham DH1 3LE, United Kingdom}
\newcommand{\ED}{University of Edinburgh, EH8 9YL Edinburgh, United Kingdom}
\newcommand{\EPFL}{Institute of Physics, Laboratory of Astrophysics, Ecole Polytechnique Fédérale de Lausanne (EPFL), Observatoire de Sauverny, 1290 Versoix, Switzerland}
\newcommand{\ETH}{ETH Zurich, Institute for Particle Physics, 8093 Zurich, Switzerland}
\newcommand{\FNAL}{Fermi National Accelerator Laboratory, Batavia, IL 60510}
\newcommand{\FQAUB}{Dept. de F\' isica Qu\` antica i Astrof\' isica, Universitat de Barcelona, Mart\' i i Franqu\` es 1, E08028 Barcelona, Spain}
\newcommand{\FSU}{Florida State University, Tallahassee, FL 32306}
\newcommand{\Glasgow}{University of Glasgow, G12 8QQ Glasgow, United Kingdom}
\newcommand{\GRAPPA}{GRAPPA Institute, University of Amsterdam, Science Park 904, 1098 XH Amsterdam, The Netherlands}
\newcommand{\GSFC}{Goddard Space Flight Center, Greenbelt, MD 20771 USA}
\newcommand{\GWU}{George Washington University, Washington, DC 20052}
\newcommand{\Hampton}{Hampton University, Hampton, VA 23668}
\newcommand{\HarvardPhys}{Department of Physics, Harvard University, Cambridge, MA 02138, USA}
\newcommand{\Haverford}{Haverford College, 370 Lancaster Ave, Haverford PA, 19041, USA}
\newcommand{\Hawaii}{University of Hawaii, Honolulu, HI 96822}
\newcommand{\HKUST}{The Hong Kong University of Science and Technology, Hong Kong SAR, China}
\newcommand{\houston}{University of Houston, Houston, TX 77204}
\newcommand{\IAP}{Institut d'Astrophysique de Paris (IAP), CNRS \& Sorbonne University, Paris, France}
\newcommand{\IAS}{Institute for Advanced Study, Princeton, NJ 08540}
\newcommand{\IBS}{Institute for Basic Science (IBS), Daejeon 34051, Korea}
\newcommand{\ICC}{ICC, University of Barcelona, IEEC-UB, Mart\' i i Franqu\` es, 1, E08028 Barcelona, Spain}
\newcommand{\ICCD}{Institute for Computational Cosmology, Department of Physics, Durham University, South Road, Durham, DH1 3LE, UK}
\newcommand{\ICE}{Institute of Space Sciences (ICE, CSIC), Campus UAB, Carrer de Can Magrans, s/n, 08193 Barcelona, Spain}
\newcommand{\ICRR}{Institute for Cosmic Ray Resaerch, The University of Tokyo, 456 Higashi-Mozumi, Kamioka, Hida, Gifu 506-1205, Japan}
\newcommand{\ICTP}{International Centre for Theoretical Physics, Strada Costiera, 11, I-34151 Trieste, Italy}
\newcommand{\IFAE}{Institut de Fisica d’Altes Energies, The Barcelona Institute of Science and Technology, Campus UAB, 08193 Bellaterra (Barcelona), Spain}
\newcommand{\IFPU}{IFPU - Institute for Fundamental Physics of the Universe, Via Beirut 2, 34014 Trieste, Italy}
\newcommand{\IFT}{Instituto de Fisica Teorica UAM/CSIC, Universidad Autonoma de Madrid, 28049 Madrid, Spain}
\newcommand{\IFUNAM}{IFUNAM - Instituto de F\'{i}sica, Universidad Nacional Aut\'onoma de M\'etico, 04510 CDMX, M\'exico}
\newcommand{\IHEP}{Institute of High Energy Physics, Austrian Academy of Sciences, 1050 Vienna, Austria}
\newcommand{\Imperial}{Theoretical Physics, Blackett Laboratory, Imperial College, London, SW7 2AZ, U.K.}
\newcommand{\Indiana}{Indiana University, Bloomington, IN 47405}
\newcommand{\INAFOATs}{INAF - Osservatorio Astronomico di Trieste, Via G.B. Tiepolo 11, 34143 Trieste, Italy}
\newcommand{\INAFOAS}{INAF - Osservatorio di Astrofisica e Scienza dello Spazio di Bologna, via Piero Gobetti 93/3, I-40129 Bologna, Italy}
\newcommand{\INFNCag}{Istituto Nazionale di Fisica Nucleare, Sezione di Cagliari,  09126 Cagliari, Italy}
\newcommand{\INFNCat}{Istituto Nazionale di Fisica Nucleare, Sezione di Catania, 95125 Catania, Italy}
\newcommand{\INFNG}{Istituto Nazionale di Fisica Nucleare, Sezione di Genova, 16146 Genova, Italy}
\newcommand{\INFN}{INFN – National Institute for Nuclear Physics, Via Valerio 2, I-34127 Trieste, Italy}
\newcommand{\INFNFE}{Istituto Nazionale di Fisica Nucleare, Sezione di Ferrara, 40122, Italy }
\newcommand{\INFNLNF}{Istituto Nazionale di Fisica Nucleare, Laboratori Nazionali di Frascati, 00044 Frascati, Italy}
\newcommand{\INFNLNS}{Istituto Nazionale di Fisica Nucleare, Laboratori Nazionali del Sud, 95125 Catania, Italy}
\newcommand{\INFNN}{Istituto Nazionale di Fisica Nucleare, Sezione di Napoli, 80125 Napoli, Italy }
\newcommand{\INFNRM}{Istituto Nazionale di Fisica Nucleare, Sezione di Roma, 00185 Roma, Italy}
\newcommand{\INFNT}{Istituto Nazionale di Fisica Nucleare, Sezione di Torino, 10125, Italy }
\newcommand{\ioa}{Institute of Astronomy, University of Cambridge,Cambridge CB3 0HA, UK}
\newcommand{\IPP}{Institute for Particle Physics, BC V8W 3P6 Victoria, Canada}
\newcommand{\IPMU}{Kavli Insitute for the Physics and Mathematics of the Universe (WPI), University of Tokyo, 277-8583 Kashiwa , Japan}
\newcommand{\IPNL}{Universit\'e de Lyon, F-69622, Lyon, France; Universit\'e de Lyon 1, Villeurbanne; CNRS/IN2P3, Institut de Physique Nucl\'eaire de Lyon}
\newcommand{\IRFU}{IRFU, CEA, Universit\'e Paris-Saclay, F-91191 Gif-sur-Yvette, France}
\newcommand{\ITFA}{Institute for Theoretical Physics, University of Amsterdam, Science Park 904, 1098 XH Amsterdam, The Netherlands}
\newcommand{\IUCAA}{The Inter-University Centre for Astronomy and Astrophysics, Pune, 411007, India}
\newcommand{\Jerusalem}{Hebrew University of Jerusalem, 91904 Jerusalem, Israel}
\newcommand{\JHU}{Johns Hopkins University, Baltimore, MD 21218}
\newcommand{\JLAB}{Thomas Jefferson National Laboratory, Newport News, VA 23606}
\newcommand{\JPL}{Jet Propulsion Laboratory, California Institute of Technology, Pasadena, CA, USA}
\newcommand{\KASSI}{Korea Astronomy and Space Science Institute, Daejeon 34055, Korea}
\newcommand{\kavli}{Kavli Institute for Cosmology, Cambridge, UK, CB3 0HA}
\newcommand{\KIAS}{School of Physics, Korea Institute for Advanced Study, 85 Hoegiro, Dongdaemun-gu, Seoul 130-722, Korea}
\newcommand{\KICP}{Kavli Institute for Cosmological Physics, Chicago, IL 60637}
\newcommand{\KIPAC}{Kavli Institute for Particle Astrophysics and Cosmology, Stanford 94305}
\newcommand{\KINGS}{King's College London, WC2R 2LS London, United Kingdom}
\newcommand{\Kobe}{Kobe University, 657-8501 Kobe, Japan}
\newcommand{\KPH}{Johannes Gutenberg University, 55128 Mainz, Germany}
\newcommand{\KPMU}{University of Tokyo, 277-8583  Kashiwa , Japan}
\newcommand{\KSU}{Kansas State University, Manhattan, KS 66506}
\newcommand{\Lafayette}{Lafayette College, Easton, PA 18042}
\newcommand{\LANL}{Los Alamos National Laboratory, Los Alamos, NM 87545}
\newcommand{\LBL}{Lawrence Berkeley National Laboratory, Berkeley, CA 94720}
\newcommand{\Leiden}{Lorentz Institute, Leiden University, Niels Bohrweg 2,Leiden, NL 2333 CA, The Netherlands}
\newcommand{\Liverpool}{University of Liverpool,  L69 7ZE Liverpool , United Kingdom}
\newcommand{\LLNL}{Lawrence Livermore National Laboratory, Livermore, CA, 94550}
\newcommand{\LPC}{Universit\'e Clermont Auvergne, CNRS/IN2P3, Laboratoire de Physique de Clermont, F-63000 Clermont-Ferrand, France}
\newcommand{\LPNHE}{Sorbonne Universit\'e, Universit\'e Paris Diderot, CNRS/IN2P3, Laboratoire de Physique Nucl\'eaire et de Hautes Energies, LPNHE, 4 Place Jussieu, F-75252 Paris, France}
\newcommand{\McGill}{McGill University, Montreal, QC H3A 2T8, Canada}
\newcommand{\Melbourne}{School of Physics, The University of Melbourne, Parkville, VIC 3010, Australia}
\newcommand{\Mines}{Colorado School of Mines, Golden, CO 80401}
\newcommand{\MIT}{Massachusetts Institute of Technology, Cambridge, MA 02139}
\newcommand{\MPE}{Max-Planck-Institut f\"{u}r extraterrestrische Physik (MPE), Giessenbachstrasse 1, D-85748 Garching bei M\"unchen, Germany}
\newcommand{\MPIA}{Max-Planck-Institut f\"{u}r Astrophysik, Karl-Schwarzschild-Str. 1, 85741 Garching, Germany}
\newcommand{\MPP}{Max-Planck-Institut f\"{u}r Physik (Werner-Heisenberg-Institut), F\"ohringer Ring 6, D-80805 M\"unchen, Germany}
\newcommand{\LUPM}{Laboratoire Univers et Particules de Montpellier, Univ. Montpellier and CNRS, 34090 Montpellier, France}
\newcommand{\NAOC}{National Astronomical Observatories, Chinese Academy of Sciences, PR China}
\newcommand{\NASAAmes}{NASA Ames Research Center, Moffett Field, CA 94035, USA}
\newcommand{\NCBJ}{National Center for Nuclear Research, Ul.Pasteura 7,Warsaw, Poland}
\newcommand{\NCU}{National Central University, Taoyuan City 32001, Taiwan (R.O.C.)}
\newcommand{\NCSU}{Physics Department, North Carolina State Universitym, 2401 Stinson Dr, Raleigh, NC 27607}
\newcommand{\ND}{University of Notre Dame,vNotre Dame, IN 46556}
\newcommand{\NIU}{Northern Illinois University, DeKalb, Illinois 60115}
\newcommand{\NMSU}{New Mexico State University, Las Cruces, NM 88003}
\newcommand{\NOAO}{National Optical Astronomy Observatory, 950 N. Cherry Ave., Tucson, AZ 85719 USA}
\newcommand{\Northwestern}{Northwestern University, Evanston, IL 60201}
\newcommand{\Nottingham}{University of Nottingham, NG7 2RD Nottingham, United Kingdom}
\newcommand{\NWU}{Northwestern University, Evanston, IL 60208}
\newcommand{\NYU}{New York University, New York, NY 10003}
\newcommand{\OK}{ University of Oklahoma, Norman, OK 73019}
\newcommand{\ORNL}{Oak Ridge National Laboratory, Oak Ridge, TN 37831}
\newcommand{\OSU}{The Ohio State University, Columbus, OH 43212}
\newcommand{\OU}{Department of Physics and Astronomy, Ohio University, Clippinger Labs, Athens, OH 45701, USA}
\newcommand{\OskarKlein}{Oskar Klein Centre for Cosmoparticle Physics, Stockholm University, AlbaNova, Stockholm SE-106 91, Sweden}
\newcommand{\Oxford}{The University of Oxford, Oxford OX1 3RH, UK}
\newcommand{\Oxy}{Occidental College, Los Angeles, CA 90041}
\newcommand{\ParisSud}{Universit\'{e} Paris-Sud, LAL, UMR 8607, F-91898 Orsay Cedex, France \& CNRS/IN2P3, F-91405 Orsay, France}
\newcommand{\PI}{Perimeter Institute, Waterloo, Ontario N2L 2Y5, Canada}
\newcommand{\Pitt}{University of Pittsburgh and PITT PACC, Pittsburgh, PA 15260}
\newcommand{\PNNL}{Pacific Northwest National Laboratory ,Richland, WA 99352}
\newcommand{\PNPI}{Petersburg Nuclear Physics Institute, 188300 Gatchina, Russia}
\newcommand{\Port}{Institute of Cosmology \& Gravitation, University of Portsmouth, Dennis Sciama Building, Burnaby Road, Portsmouth PO1 3FX, UK}
\newcommand{\Princeton}{Princeton University, Princeton, NJ 08544}
\newcommand{\PSU}{The Pennsylvania State University, University Park, PA 16802}
\newcommand{\Purdue}{Purdue University, West Lafayette, IN 47907}
\newcommand{\PW}{Participation Worldscope, Sedona, Arizona and Hong Kong, SAR PRC}
\newcommand{\Queens}{Queen's University , K7L 3N6 Kingston, Canada}
\newcommand{\Queensland}{The University of Queensland, School of Mathematics and Physics, QLD 4072, Australia}
\newcommand{\QMUL}{Queen Mary University of London, Mile End Road, London E1 4NS, United Kingdom}
\newcommand{\RAL}{Radio Astronomy Laboratory, University of California Berkeley, Berkeley, CA 94720, USA}
\newcommand{\Rice}{Department of Physics \& Astronomy, Rice University, Houston, Texas 77005, USA}
\newcommand{\RIT}{Rochester Institute of Technology}
\newcommand{\RomaS}{Dipartimento di Fisica, Universit\`{a} La Sapienza, P. le A. Moro 2, Roma, Italy}
\newcommand{\RUG}{Kapteyn Astronomical Institute, University of Groningen, P.O. Box 800, 9700 AV Groningen, The Netherlands}
\newcommand{\Rutgers}{Department of Physics and Astronomy, Rutgers, the State University of New Jersey, 136 Frelinghuysen Road, Piscataway, NJ 08854, USA}
\newcommand{\Sanford}{Sanford Underground Research Facility, Lead, SD 57754}
\newcommand{\Sassari}{Universit\`a di Sassari, 07100 Sassari,  Italy}
\newcommand{\SCIPP}{University of California at Santa Cruz, Santa Cruz, CA 95064}
\newcommand{\Sejong}{Department of Physics and Astronomy, Sejong University, Seoul, 143-747, Korea}
\newcommand{\Sheffield}{University of Sheffield, S3 7RH Sheffield, United Kingdom}
\newcommand{\SHAO}{Shanghai Astronomical Observatory (SHAO), Nandan Road 80, Shanghai 200030, China}
\newcommand{\Siena}{Siena College, 515 Loudon Road, Loudonville, NY 12211, USA}
\newcommand{\SISSA}{SISSA - International School for Advanced Studies, Via Bonomea 265, 34136 Trieste, Italy}
\newcommand{\SimonFraser}{Department of Physics, Simon Fraser University, Burnaby, British Columbia, Canada V5A 1S6}
\newcommand{\SLAC}{SLAC National Accelerator Laboratory, Menlo Park, CA 94025}
\newcommand{\SMU}{Southern Methodist University, Dallas, TX 75275}
\newcommand{\SNOLAB}{SNOLAB, Lively, ON P3Y 1N2, Canada}
\newcommand{\SoCal}{University of Southern California, CA 90089 }
\newcommand{\Stanford}{Stanford University, Stanford, CA 94305}
\newcommand{\StonyBrook}{Stony Brook University, Stony Brook, NY 11794}
\newcommand{\STSCI}{Space Telescope Science Institute, Baltimore, MD 21218}
\newcommand{\SUNYA}{University at Albany SUNY, Albany, NY 12222}
\newcommand{\SussexAstronomy}{Astronomy Centre, School of Mathematical and Physical Sciences, University of Sussex, Brighton BN1 9QH, United Kingdom}
\newcommand{\Syracuse}{Syracuse University, Syracuse, NY 13244}
\newcommand{\Tamu}{Texas AandM University, College Station, TX 77843 }
\newcommand{\Techsource}{Techsource Incorporated, Los Alamos, NM 87544}
\newcommand{\TelAviv}{Tel-Aviv University,  69978 Tel-Aviv, Israel}
\newcommand{\Temple}{Temple University, Philadelphia, PA 19122}
\newcommand{\TIFR}{Tata Institute of Fundamental Research, Homi Bhabha Road, Mumbai 400005 India}
\newcommand{\Tsinghua}{Department of Physics and Tsinghua Center for Astrophysics, Tsinghua University, Beijing 100084, P R China}
\newcommand{\TUM}{Technical University of Munich,  80333 Munich, Germany}
\newcommand{\UA}{University of Alabama, Tuscaloosa, AL 35487}
\newcommand{\UAS}{Department of Astronomy/Steward Observatory, University of Arizona, Tucson, AZ  85721}
\newcommand{\UAM}{Universidad Aut\'onoma de Madrid, 28049, Madrid, Spain}
\newcommand{\UBC}{University of British Columbia, Vancouver, BC V6T 1Z1, Canada}
\newcommand{\UCB}{Department of Astronomy, University of California Berkeley, Berkeley, CA 94720, USA}
\newcommand{\UCBP}{Department of Physics, University of California Berkeley, Berkeley, CA 94720, USA}
\newcommand{\UCBSSL}{Space Sciences Laboratory, University of California Berkeley, Berkeley, CA 94720, USA}
\newcommand{\UCD}{University of California at Davis, Davis, CA 95616}
\newcommand{\UChicago}{University of Chicago, Chicago, IL 60637}
\newcommand{\UCI}{University of California, Irvine, CA 92697}
\newcommand{\UCLA}{University of California at Los Angeles, Los Angeles,  CA 90095}
\newcommand{\UCL}{University College London, WC1E 6BT London, United Kingdom}
\newcommand{\UCR}{University of California at Riverside, Riverside, CA 92521}
\newcommand{\UCSB}{University of California at Santa Barbara, Santa Barbara, CA 93106}
\newcommand{\UCSC}{University of California at Santa Cruz, Santa Cruz, CA 95064}
\newcommand{\UCSD}{University of California San Diego, La Jolla, CA 92093}
\newcommand{\UFL}{University of Florida, Gainesville, FL 32611}
\newcommand{\UFN}{Universit\`a Federico II di Napoli, 80125 Napoli, Italy}
\newcommand{\UGTO}{Divisi\'on de Ciencias e Ingenier\'ias, Universidad de Guanajuato, Le\'on 37150, M\'exico}
\newcommand{\UKY}{University of Kentucky, Lexington, KY 40506}
\newcommand{\UMD}{University of Maryland, College Park, MD 20742
\newcommand{\UMiami}{University of Miami, Coral Gables, FL 33124}}
\newcommand{\UMich}{University of Michigan, Ann Arbor, MI 48109}
\newcommand{\UMN}{University of Minnesota, Minneapolis, MN 55455}
\newcommand{\UnB}{Instituto de F\'{i}sica, Universidade de Bras\'{i}lia, 70919-970, Bras\'{i}lia, DF, Brazil}
\newcommand{\UNC}{University of North Carolina at Chapel Hill, Chapel Hill, NC 27599}
\newcommand{\UNH}{University of New Hampshire, Durham, NH 03824}
\newcommand{\UNIMI}{Dipartimento di Fisica ``Aldo Pontremoli'', Universit\`a{} degli Studi di Milano, via Celoria 16, 20133 Milano, Italy}
\newcommand{\UNIPD}{Dipartimento di Fisica e Astronomia ``G. Galilei'',Universit\`a degli Studi di Padova, via Marzolo 8, I-35131, Padova, Italy}
\newcommand{\UNM}{University of New Mexico, Albuquerque, NM 87131}
\newcommand{\UNV}{University of Nevada, Reno, NV 89557}
\newcommand{\UoM}{Jodrell Bank Center for Astrophysics, School of Physics and Astronomy, University of Manchester, Oxford Road, Manchester, M13 9PL, UK}
\newcommand{\UPenn}{Department of Physics and Astronomy, University of Pennsylvania, Philadelphia, Pennsylvania 19104, USA}
\newcommand{\UR}{Department of Physics and Astronomy, University of Rochester, 500 Joseph C. Wilson Boulevard, Rochester, NY 14627, USA}
\newcommand{\UrbanaC}{Department of Physics, University of Illinois at Urbana-Champaign, Urbana, Illinois 61801, USA}
\newcommand{\USC}{The University of South Carolina, Columbia, SC 29208}
\newcommand{\USD}{The University of South Dakota, Vermillion, SD 57069}
\newcommand{\UTD}{University of Texas at Dallas, Texas 75080}
\newcommand{\Utenn}{The University of Tennessee, Knoxville, TN 37996}
\newcommand{\Utah}{University of Utah, Department of Physics and Astronomy, 115 S 1400 E, Salt Lake City, UT 84112, USA}
\newcommand{\UVA}{University of Virginia, Charlottesville, VA 22903}
\newcommand{\Uvic}{University of Victoria, BC V8P 5C2 Victoria, Canada}
\newcommand{\UWaterloo}{Department of Physics and Astronomy, University of Waterloo, 200 University Ave W, Waterloo, ON N2L 3G1, Canada}
\newcommand{\UWMadison}{Department of Physics, University of Wisconsin - Madison, Madison, WI 53706}
\newcommand{\UW}{University of Washington, Seattle 98195}
\newcommand{\UWC}{Department of Physics \& Astronomy, University of the Western Cape, Cape Town 7535, South Africa}
\newcommand{\Vanderbilt}{Physics \& Astronomy Department, Vanderbilt University, PMB 401807, 2301 Vanderbilt Place, Nashville, TN 37235}
\newcommand{\VSI}{Van Swinderen Institute for Particle Physics and Gravity, University of Groningen, Nijenborgh 4, 9747~AG~Groningen, The~Netherlands}
\newcommand{\VT}{Virginia Tech, Blacksburg, VA 24061}
\newcommand{\VUU}{Virginia Union University, Richmond, Virginia, 23220}
\newcommand{\WCA}{Centre for Astrophysics, University of Waterloo, Waterloo, Ontario N2L 3G1, Canada}
\newcommand{\Weizmann}{Weizmann Institute of Science, 76100 Rehovot, Israel}
\newcommand{\Wellesley}{Wellesley College, Wellesley, MA 02481}
\newcommand{\wiscIce}{University of Wisconsin, Madison, WI 53706}
\newcommand{\WM}{College of William and Mary, Newport News, VA 23606}
\newcommand{\WUSL}{Washington University in St Louis, St. Louis, MO 63130}
\newcommand{\WVU}{CSEE, West Virginia University, Morgantown, WV 26505, USA}
\newcommand{\WVUGWAC}{Center for Gravitational Waves and Cosmology, West Virginia University, Morgantown, WV 26505, USA}
\newcommand{\Wyoming}{Department of Physics and Astronomy, University of Wyoming, Laramie, WY 82071, USA}
\newcommand{\Yale}{Department of Physics, Yale University, New Haven, CT 06520}
\newcommand{\YorkU}{Department of Physics and Astronomy, York University, Toronto, Ontario M3J 1P3, Canada}

\textbf{Co-authors / Endorsers:} Kevork N.\ Abazajian$^{1}$, 
Zeeshan Ahmed$^{2}$, 
David Alonso$^{3}$, 
Mustafa A. Amin$^{4}$, 
Behzad Ansarinejad$^{5}$, 
Robert Armstrong$^{6}$, 
Jacobo Asorey$^{7}$, 
Arturo Avelino$^{8}$, 
Santiago Avila$^{9}$, 
Carlo Baccigalupi$^{10,11,12}$
% ...


\begin{raggedright}
\setlength{\columnsep}{18pt}
\begin{multicols}{2}

 $^{1}$ \UCI \\
$^{2}$ \SLAC \\
$^{3}$ \Oxford \\
$^{4}$ \Rice \\
$^{5}$ \Durham \\
$^{6}$ \LLNL \\
$^{7}$ \KASSI \\
$^{8}$ \CfA \\
$^{9}$ \UAM \\
$^{10}$ \SISSA \\
$^{11}$ \IFPU \\
$^{12}$ \INFN \\
% ...

\normalsize
\end{multicols}
\end{raggedright}

\pagebreak
\pagenumbering{arabic}

\setlength{\parskip}{3pt}
\setlength{\parindent}{18pt}
\section{Introduction}
\vspace*{-9pt}

Twenty years since the discovery of cosmic acceleration, the laws of physics on the largest scales remain an enigma.  The phenomenon continues to 
drive us to consider bold transformations in our understanding of cosmic
forces.
Broadly speaking, theories of dark
energy fall into three major classes. The first and simplest class
is the cosmological constant, which fits all the available data
but which is theoretically unsatisfactory because it is fine-tuned and
sensitive to the UV completion of gravity. This implies that any fundamental
theory that could give rise to a stable value must have some new
physics associated with it \cite{Padilla:2015aaa,Burgess:2013ara}.

The second class of models includes quintessence-type theories where
dark energy is a component with novel properties, typically explained
in terms of a scalar field. These  models of dark energy
are the simplest dynamical models of dark energy and are a low-redshift
equivalent of inflation\cite{Carroll2001}. They have recently been
discussed within a somewhat controversial debate about the swampland
conjecture, which states that theories that can produce a slowly
rolling $w\simeq -1$ universe like the one we observe do not have
consistent UV completions within string theory (\cite{2018arXiv181210448R},
but see also \cite{2019arXiv190110489H}). This conjecture severely
constrains the space of string-theory compatible models \cite{Agrawal:2018own,Heisenberg:2018yae}.

The third class of models includes modified-gravity theories in which cosmic acceleration
is caused by some extension of the gravitational sector. These
theories have been significantly constrained after the discovery
of gravitational wave source GW170817 for which the optical counterpart
has been discovered, which constrained the speed of tensor mode propagation
$c_t=c$ with a fractional accuracy of $10^{-15}$. Nevertheless, there
are significant caveats: i) the cosmic acceleration and the LIGO event
are separated by twenty orders of magnitude in energy scales and
the effective theories can be very different \cite{deRham:2018red},
and ii) modified-gravity theories can predict significant time
dependence, so a single event cannot completely constrain the
underlying physics. Significant parts of parameter space thus
remain open. Finally, there are somewhat more exotic theoretical
paths towards generating accelerated expansion, like novel coupling
between baryons and dark matter \cite{2017PhRvD..95l3530B} or
involving neutrino physics \cite{2004JCAP...10..005F,2017arXiv170204244M}.

The main observables that will help us distinguish between these scenarios are the precise characterization of the \emph{expansion history}, the \emph{gravitational slip} ($\eta=\Phi/\Psi$, where $\Phi$ and $\Psi$ are gravitational potentials in the time and space perturbations of the metric) and the \emph{effective large-scale Newton's constant} $\Geff$ \cite{2018arXiv181203969S}. Both $\eta$ and $\Geff$ are unity in the standard GR, but can be time- and scale-dependent quantities in modified-gravity theories. This has led to renewed interest in measuring the growth of structure and the combination of datasets with the explicit purpose of constraining these two parameters \cite{2010Natur.464..256R, 2012PhR...513....1C,2013PhRvD..87b3501A,Huterer:2013xky,Alonso:2016suf,2018MNRAS.480.3725S,2019MNRAS.482.3274R,2019arXiv190103686S,Ishak2019} 

Faced with this compelling mystery, astrophysicists have mounted 
an ambitious multi-faceted campaign to study the behavior of the
Universe on the large scales.  This program has been undeniably 
successful: data quality has improved radically over the past 
decade, with new leaps expected early in the 2020s from the upcoming
facilities.  There have been major improvements in parameter 
constraints: for a simple equation of state $w(z)=p/\rho c^2$,
we have measured a constant $w$ to be consistent with the cosmological
constant ($w=-1$) to about $\pm$0.05.  And while there is
consistency in many aspects of the results, there are also active 
disagreements in other parts, such as the value of the Hubble
constant and perhaps the amplitude of late-time matter clustering \cite{Tanabashi:2018oca, 2018ApJ...855..136R,2018MNRAS.476.4662V,2018PhRvD..98d3526A,2018arXiv180706209P,2018arXiv180909148H}.


\section{Entering the decade of precision dark energy science}
\label{sec:2}
\vspace*{-9pt}

There is a rich portfolio of 
observational methods that we expect will drive the study of cosmic 
acceleration in the coming decade.  We stress that the whole
is more than the sum of the parts.  The methods reinforce each other
both in terms of statistical leverage and control of systematic uncertainties.
Cross-correlations and data combination have emerged as indispensable
tools for both controlling systematic uncertainties and isolating
particularly informative aspects of theories.  
During this decade, use of blinded analysis has become 
the norm for most measurements of cosmic acceleration, so as to avoid confirmation
bias; developing methods for blinded analysis that will work for
surveys at the next level of precision will be important as the
field moves forward.  For all of these methods, marginalizing over
systematic uncertainties has resulted in expanded parameter spaces,
and the field continues to work on building and validating models for major systematic
uncertainties that will work at the level needed for upcoming surveys.

There are two main classes of methods to study dark energy.  The
first measures the \emph{expansion history}, particularly through the
study of the distance-redshift relation.  
The second class
measures the \emph{growth} of matter density fluctuations, which
is impacted by the large-scale gravitational forces.  Because dark
energy is typically smooth, it slows the growth of fluctuations and
decreases the number of dark matter halos of a given mass.  However,
growth measurements offer more than an increase in statistical
precision; they provide an important consistency check.  In a
broad class of quintessence theories, the expansion history predicts
the behavior of  the growth of fluctuations, so any evidence for
inconsistency there would necessarily imply some non-standard physics
in the gravitational sector.

We now briefly summarize what we see as the major methods for the coming decade.

\newcommand{\method}[5]{
\smallskip \noindent \textbf{#1:}\ 
\underline{\smash{Description:}} #2 
\underline{\smash{Status:}} #3
\underline{\smash{Future challenges:}} #4
\underline{\smash{Unique selling points:}} #5\\
}

\medskip
\noindent {\large \textbf{Expansion history}}
\linebreak
\method{Baryonic Acoustic Oscillations (BAOs)}{
The BAO peak is a feature in the correlation function of a tracer of large-scale structure, acting as a standard ruler and thus allowing measurements of distances and expansion rates as a function of redshift.}{
Numerous experiments have measured BAOs with high precision in the past decade, including 2dFGRS, 6dFGS, WiggleZ, SDSS II, BOSS and eBOSS, using both galaxies and the Lyman-$\alpha$ forest as tracers. In the 2020s, DESI, PFS, and Euclid will carry out high precision galaxy BAO measurements to $z\sim 2$ with DESI Lyman-$\alpha$, HETDEX, and WFIRST galaxy surveys reaching $z\sim 3$.} {
The main future challenge lies in obtaining sufficiently large spectroscopic samples at ever increasing volumes. At redshifts beyond $2$, non-galaxy tracers such as the Lyman-$\alpha$ forest and 21\,cm could be optimal.}{
BAO is arguably the most mature method and is theoretically and experimentally well understood.}

\method{Supernovae Type Ia (SNe Ia)}
{
Type Ia supernovae (SNe Ia) are bright standard candles that probe the expansion history of the Universe through calibrating the luminosity distance as a function of redshift.
}{
The CfA Supernova program,  Carnegie Supernova Program (CSP), SDSS-II SN survey, Supernova Legacy Survey (SNLS), PanSTARRS, DES Supernova program, ESSENCE, GOODS survey, CANDELS/CLASH, Supernova Cosmology Project (SCP) and others have measured SNe Ia over wide range of redshifts. In the 2020's, LSST will deliver $O(10^5)$  photometric SNe Ia, WFIRST will measure SNe Ia in the infrared with the resolution and
photometric stability achievable from space, while the Foundation survey will yield many low-redshift SNe.
}{
Photometric SNe require very precise photometric  and filter calibrations, spectroscopic characterization and updated light curve models. 
}{
SN are the dominant probe of the Hubble diagram at the lowest redshifts, $z<0.5$, but are also bright enough to be used over a wide range of redshift.}

\method{Time delays strong lensing}{
Time delays measured between the multiple images of the same object provide measurements of the Hubble parameter independently of other distance measures.}{
Dedicated observational programs like H0LiCOW/COSMOGRAIL and STRIDES, which combine ground and space observations of 20 lensed quasar systems, are publishing competitive $H_0$ constraints.  Future surveys like
LSST will discover orders of magnitude more systems, with $\sim$400
expected to be suitable for dark energy science, and measure hundreds of strongly gravitationally-lensed supernovae, which should enable similar measurement with less monitoring. 
}{Mass modeling, external convergence, and correlations between the modeling and the cosmological parameters remain the largest systematic uncertainty.}{
This technique is independent from other distance indicators and can achieve precision to constrain dark-energy parameters from a relatively small number of systems.}

\method{Standard Sirens}{
Gravitational waves (GW) from the inspiral of two massive objects are a powerful measure of a source's luminosity distance.}{
The discovery of GWs by Advanced LIGO in 2016 ushered in the era of gravitational wave astrophysics. From one source with an identified optical counterpart, a 7\% measurement of the Hubble constant was obtained. The 2020s could see standard sirens providing a 2\% determination of $H_0.$}{
As more sources are discovered, selection effects in both the GW surveys and the EM follow-up programs will need to be included in systematic error analyses.}{
The amplitude of the standard siren signal is computed from fundamental
physics and does not rely on empirical calibration. }

\medskip
\noindent {\large \textbf{Growth}}
\linebreak
\method{Weak Gravitational Lensing}{
Measurements of coherent distortions in galaxy shapes due to weak gravitational lensing reveal the distribution of dark matter in the Universe.
}{
Dedicated surveys including CFHTLS, KIDS, DES and HSC have achieved statistical precision of a few percent in the amplitude of matter fluctuations at redshift $z \lesssim 1.2$.  Lensing surveys in the 2020s from the ground (LSST) and space (WFIRST and Euclid) will cover large sky areas at significant depths. Dedicated space-based observations will enable major advances via high resolution, wavefront stability, and access to the NIR.
}{
  Photometric redshifts and blending will require further methodological improvements. Another challenge is the theoretical modeling of the signal in the presence of astrophysical systematics (intrinsic alignments and baryonic effects).
}{
Combining tomographic measurements of weak lensing with measurements of the expansion history may be the most effective way to probe GR potentials and distinguish between dark energy and modified gravity.
}

\method{Cosmic Microwave Background (CMB) lensing}{
Measuring distortions in the CMB fluctuations can probe weak gravitational lensing to the surface of last scattering.
}{
Lensing reconstructions from the Planck satellite currently provide the highest signal-to-noise measurements on  the linear amplitude of fluctuations at $z<6$. Simons Observatory, CMB-S4 and PICO will improve these limits by an order of magnitude in the coming decades.
}{
In order to minimize systematic uncertainties from secondary anisotropies, the CMB lensing will rely on the polarization signal, which is weaker and has its own, yet to be fully understood foregrounds. 
}{
CMB lensing provides a long redshift lever-arm and has fewer observational systematic uncertainties compared to galaxy lensing.
}
\method{Peculiar Velocities}{
Peculiar velocities are motions of galaxies not comoving with the expansion of the Universe. They can be measured for individual objects using redshift and a distance indicator.}{
6dFGS and 2MTF have measured tens of thousands of galaxy peculiar velocities with accuracies of  20\%. Supernova surveys have higher precision, but are limited by low numbers.  Upcoming surveys like TAIPAN, WALLABY+WNSHS, ZTF and LSST will make large peculiar-velocity catalogs,  enabling tight constraints on growth at low redshift.}{
Proper handling of the asymmetric uncertainties on distance indicators is crucial.}{
Understanding the local peculiar velocity field constrains dark energy and dark matter directly, and helps with systematic control in other probes by characterizing the local density environment.}

\method{Redshift-space distortions (RSDs)}{
  Redshift-space distortions are peculiar velocities detected statistically as  an apparent anisotropy of the measured correlations in any large-scale structure survey.}{
Spectroscopic galaxy surveys, including 6dFGS, WiggleZ, VIPERS, BOSS and eBOSS, have measured the growth parameter $f\sigma_8$ with 3-10\% precision depending on modeling assumptions. In the coming decade DESI, PFS, Euclid and WFIRST will make percent-level measurements at $z<1.8$ and WFIRST will push to $z\simeq 3$.
}{
Theoretical modeling of non-linear effects, and the connection between light and mass, remain the main issue.
}{
One of the most direct ways of measuring the growth rate with the potential to significantly improve signal-to-noise with better modelling. 
}

\method{Kinetic Sunyaev-Zeldovich (kSZ) effect}{
The kSZ effect is the Doppler shift of CMB photons caused by scattering off the plasma in late-time galaxies and clusters.
}{
The first detection of the pairwise kSZ was made in 2012 using data from the  Atacama Cosmology Telescope and BOSS galaxy survey. This science will benefit from the large survey area CMB experiments (SO, CMB-S4, PICO), and their cross-correlation with optical galaxy surveys (LSST, DESI). 
}{
A difficulty in constraining growth using kSZ measurements is the degeneracy with the optical depth in galaxy clusters and groups.}{
This is an independent probe of the velocity field at low redshift, with different systematics and modelling assumption compared to redshift-space distortions.}

\method{Galaxy Clusters}{
Galaxy clusters are the most massive, gravitationally bound structures in the Universe, and their abundance provides a sensitive probe of growth.
}{
 Planned optical/IR (LSST, WFIRST, Euclid), Sunyaev-Zeldovich (SO, CMB-S4, PICO), and X-ray (eROSITA, ATHENA) surveys will provide cluster catalogs over a wide range in mass and out to unprecedentedly high redshifts.
}{
  The main difficulty is obtaining an accurate absolute cluster mass calibration and  precise relative mass estimates. The latter are considerably improved using X-ray data; while eROSITA will provide these at relatively low redshifts, an ongoing source of high-throughput, targeted X-ray observations will be required to fully exploit the high-redshift catalogs provided by thermal SZ surveys.
}{
Galaxy clusters are a statistically sensitive probe of growth with
largely independent systematics.
}

\vspace{-15pt}
\section{Conclusions and Outlook}
\label{sec:conclusions}
\vspace*{-9pt}

The coming decade will be an exciting one for dark energy studies, 
as a new generation of powerful observational facilities comes to 
fruition.  The combination of high-precision data with growth in 
theoretical models and statistical techniques will allow a great
leap in our cosmological leverage, testing our theories in 
unprecedented ways and perhaps sharpening the fault lines in 
present results.  As we look toward the coming decade in cosmology and the study
of dark energy and modified gravity, we want to highlight the 
following themes.


\textbf{Improving statistical and systematic precision on the equation of state is essential}. The current statistical precision for the $w$CDM model is around 5\% (1 $\sigma$). To formally distinguish between a $w=-1$ model and a $-1<w<0$ model at 1:100 statistical odds, one would need to achieve sub-percent level precision. Even more importantly, dark energy models with dynamical equations of state remain significantly underconstrained. Understanding dark energy to the percent level in the acceleration era and tens of percent in the high-redshift pre-acceleration era remains one of the long-term programmatic goals of cosmology. This also requires support for further methodological  advances to reduce systematic uncertainties.

\textbf{Multiple methods bring robustness.} Characterizing dark energy and modified gravity through as many different methods as possible provides valuable cross-checks and data consistency tests as methods hit systematic floors. 
We see critical opportunities here both in tests of expansion history
(e.g., the current tension in the value of $H_0$) and growth
(e.g., the current concerns within lensing and cluster analysis 
regarding the value of $\sigma_8$).

\textbf{Multiple observatories bring robustness.} If there is evidence of deviations from General Relativity or evidence for dynamical dark energy, it is essential to cross-check results with independent experiments using multiple techniques with careful control of systematic errors.

\textbf{Cross-correlations are ever more important.} Applying similar methods over the same volume brings about numerous cross-correlations that have proven to be very valuable.  
In order to make maximal use of cross-correlations, it is essential to support simulation and data processing/analysis tools that are compatible across surveys and collaborations.

\textbf{Blind analysis is desirable but challenging.} This is especially true with upcoming complex analyses that involve numerous, often subjective, analysis choices.  
Executing a blind analyses requires careful methodological planning, extensive
support from simulations, and delicate coordination, particularly
when combining numerous methods across a broad collaboration.

\textbf{Studies of dark energy and modified gravity are related.} All but the simplest dark energy models predict modifications of gravity, although the two can be distinguished by comparing probes of the background expansion and the growth of structure \cite{Huterer:2013xky}. Both exhibit deviations from $\Lambda$CDM on cosmological scales (e.g. \cite{Alonso:2016suf,2019arXiv190201407B}) that can be tested with the same probes. They should be studied as one field.


\textbf{Dark energy science in the 2030s will require technical R\&D support.}
 The path forward into the 2030s will require an ongoing investment at the observational frontier. Whether by mapping of huge cosmic volumes or by discovery and  characterization of rare transients, improving our view of dark energy will require continued technological ambition. Design and development of a broad technical portfolio in this decade will be needed to achieve the necessary capabilities, both statistically and systematically, in a cost-effective manner.

The field of cosmology has
been adept at unifying large teams to produce and optimize
state-of-the-art facilities, the products of which have advanced
many areas of astrophysics.  We believe that the mystery of 
dark energy and the diverse range of measurements that bear on
it remains a compelling driver to motivate this development 
in the coming decade.



\pagebreak

\bibliographystyle{unsrt}
\bibliography{main}

\end{document}